# Manipulation of polar vortex chirality in oxide superlattices

Pan Chen[1,2,11], Congbing Tan[3,11], Zhexin Jiang[4,11], Peng Gao[2,5*], Yuanwei Sun[2], Xiaomei Li[1], Ruixue Zhu[2], Lei Liao[1,6], Xu Hou[4], Lifen Wang[1], Ke Qu[2], Ning Li[2], Xiaomin Li[1], Zhi Xu[1,7], Kaihui Liu[5, 8], Wenlong Wang[1,7], Jinbin Wang[9], Xiaoping Ouyang[9], Xiangli Zhong[9*], Jie Wang[4,10*] and Xuedong Bai[1,6,7*]

[1]*Beijing National Laboratory for Condensed Matter Physics, Institute of Physics, Chinese Academy of Sciences, Beijing 100190, China.*

[2]*International Center for Quantum Materials, and Electron microscopy laboratory, School of Physics, Peking University, Beijing 100871, China.*

[3]*Hunan Provincial Key Laboratory of Intelligent Sensors and Advanced Sensor Materials, School of Physics and Electronics, Hunan University of Science and Technology, 411201, Hunan Xiangtan, China.*

[4]*Department of Engineering Mechanics, Zhejiang University, 310027 Hangzhou, China.*

[5]*Collaborative Innovation Center of Quantum Matter, Beijing 100871, China.*

[6]*School of Physical Sciences, University of Chinese Academy of Sciences, Beijing 100190, China.*

[7]*Songshan Lake Materials Laboratory, Dongguan, Guangdong 523808, China.*

[8]*State Key Laboratory for Artificial Microstructure & Mesoscopic Physics, School of Physics, Peking University, Beijing 100871, China.*

[9]*School of Materials Science and Engineering, Xiangtan University, Hunan, Xiangtan 411105, China.*

[10]*Key Laboratory of Soft Machines and Smart Devices of Zhejiang Province, Zhejiang University, 310027 Hangzhou, China.*

[11]*These authors contributed equally to this work*

*E-mails: p-gao@pku.edu.cn; xlzhong@xtu.edu.cn; jw@zju.edu.cn, xdbai@iphy.ac.cn

**Abstract**

Topological polar vortices that are the electric analogues of magnetic objects, present great potential in applications of future nanoelectronics due to their nanometer size, anomalous dielectric response, and chirality. To enable the functionalities, it is prerequisite to manipulate the polar states and chirality by using external stimuli. Here, we probe the evolutions of polar state and chirality of polar vortices in $PbTiO_3$/$SrTiO_3$ superlattices under electric field by using atomically resolved in situ scanning transmission electron microscopy and phase-field simulations. We find that the adjacent clockwise and counterclockwise vortex usually have opposite chirality. The phase-field simulations suggest that the rotation reversal or axial polarization switching can lead to the chirality change. Guided by which, we experimentally validate that the vortex rotation direction can be changed by applying and subsequently removing of electric fields, offering a potential strategy to manipulate the vortex chirality. The revealed details of dynamic behavior for individual polar vortices at atomic scale and the proposed strategy for chirality manipulation provide fundamentals for future device applications.

**Introduction**

Polar vortex, a topological structure with electric dipoles continuously rotating around a stable core, exhibits new emergent ordering and properties[1-3] and thus attracts extensive research attention[4-7]. Such vortices have been theoretically predicted and experimentally realized in low dimensional ferroelectrics[8-10]. Recently, the vortex array was discovered in $(PbTiO_3)_n/(SrTiO_3)_m$ superlattices (PTO/STO, where n and m are the number of unit cells for PTO and STO, respectively) with alternating clockwise and counterclockwise rotations of electric dipoles[1]. Their small size makes them ideal storage units for high density memory applications[2, 4, 11] and the anomalous dielectric behavior or negative capacitance, has great potential for nanoelectronics to overcome the Boltzmann tyranny[4]. Moreover, these vortices carry chirality[12], providing a new order parameter for these complex oxides and thus enabling possible novel functionalities and devices by using it.

In order to enable these applications of the polar vortex, it is prerequisite to manipulate the ferroelectric order parameters and/or the vortex chirality. The theoretical investigations[13, 14] have predicted the evolution path from the vortex to trivial ferroelectrics under external electric fields, i.e., clockwise and counter-clockwise vortices melt to form *a*/c domain. Experimentally, Damodaran et al.[6] has demonstrated the interconversion between vortex and ferroelectric $a_1/a_2$ domains under electric field by atomic force microscopy (AFM). Using in situ transmission electron microscopy (TEM), Du et al.[15] showed a local transformation from vortex into dipole wave and *c* domain by applying non-contact bias and the transitioned states sustained even with the removal of electric bias, while Nelson et al.[16] have found the vortex response to the applied electric fields through a reversible transition to *a/c* domain or *a*-domain. Besides the electrical field driven phase transition, Chen et al[17]. observed mechanical stress can also cause similar phase transition between topological polar state and trivial *a*-domain. Recently, Li et al observed a ultrafast collective polarization dynamics driven by terahertz-field electrical field[18].

Despite of great progress on switching pathways and polarization dynamics for vortex, the chirality manipulation of polar vortex has been rarely explored in experiments. The chirality of polar vortex is determined by both the rotation direction and the axial-component of polarization along the vortex tube direction[12]. Theoretically, the switching of vortex rotation is predicted either by a curled electric field[19, 20] or by deliberately designed geometry of nanostructures (such as notched nanodots, asymmetric nanorings and void nanoplatelet)[21-23], which, however, is somewhat impractical for to current technologies due to the difficulties in control of such a distribution of electric fields or fabrication of these complicated nanostructures. Experimentally, characterization of chirality in polar vortex, however, is extremely difficult. To date, the transmission electron microscopy is the only approach to directly probe the rotation direction and radial-component polarization for individual polar vortex[2], while the axial-component polarization has never been directly characterized in experimental studies, not to mention tracking their evolution under external stimuli.

Here, by combining the phase field simulations and atomically resolved in situ scanning transmission electron microscopy (STEM) methods, we probe the dynamic evolutions and chirality of individual polar vortices in $PbTiO_3/SrTiO_3$ superlattices under electric field. We find that the adjacent clockwise and counterclockwise vortices usually have opposite chirality. We directly observe the rotation of chiral vortex can be changed during switching and back-switching events under simple electrical fields. Based on our phase field simulations, the rotation reversal is accompanied by the chirality reversal. Besides, we also reveal the detailed switching pathways and find that the vortex core can move laterally into *a/c* domain stripes and then become a mono non-chiral *c*-domain, which is in excellent agreements with phase field simulations. Our work reveals the reversible switching pathways of individual vortices at atomic scale and demonstrates the ability to electrically tune the vortex rotation, which may open a new strategy to control the chirality and further utilize it as a new order parameter for novel device applications.

**Results**

   **Atomic scale characterization of polar vortex.** The $(PTO)_n/(STO)_m$ thin films were synthesized on $(001)_{pc}$ (pc denotes the pseudocubic indices) $DyScO_3$ substrate with $SrRuO_3$ as the bottom electrode by pulsed-laser deposition (Methods). Figure 1a, b are low magnification dark-field images of plane-view and cross-section samples, respectively. The stripes in Fig. 1a are vortex tubes (Supplementary Fig. 1) and the alternating bright and dark contrast in dark-filed TEM image Fig. 1b correspond to the clockwise and counter-clockwise vortex array that exhibit the character of long-range ordering, wherein the out-of-plane g-vector $(002)_{pc}$ is selected to distinguish $P_{[001]}\uparrow$ and $P_{[001]}\downarrow$ components. The phase-filed simulations in Fig. 1c shows the distribution of polarization vectors corresponding to the contrast in the dark-field TEM image[24]. Thus, the core position of vortex can be located in the boundary between the dark and bright area from dark-field TEM image. The PTO/STO superlattice film shows sharp interfaces confirmed by the atomically resolved EDS mapping in Supplementary Fig. 2. Dipole configuration can be evidenced by calculating the offset of Pb/Sr and Ti sublattice on the basis of the atomically resolved HAADF-STEM image shown in Fig. 1d for the cross sectional view and plane view, from which the adjacent clockwise and counter-clockwise vortex have opposite rotation but same axial component, leading to opposite chirality. Phase-field simulations reproduce the same polarization configuration shown in Fig. 1e. Note that along the tube direction (~$[010]_{pc}$) the polarization is much smaller than that along the other directions (~$[001]_{pc}$) (Supplementary Fig. 3). Such small values are difficult to extract from quantitative analysis of electron microscopy images[25], which explains why previous electron microscopy studies rarely discussed the chirality[15, 17, 26]. It is even harder to detect the axial polarization component for the in situ TEM experiments that generally suffer relatively large vibrations and instabilities due to the introduction of the contacting probe and applied fields. Thus, we performed the in situ experiments from the cross-section view operated under lower magnification dark-field and atomically resolved HAADF modes to track the evolution of rotation direction. Furthermore, we extract the displacement maps (Fig. 1g, h) and out-of-plane lattice constant (Fig. 1i) from the atomically resolved HAADF image (Fig. 1d), which is schematically illustrated in Fig. 1f. From these maps, for each unit cell the core positions and possible domain wall positions can be located and the rotation direction can be identified[27]. Particularly, the vortex core lies in the boundary of the upward and downward out-of-plane displacement regions (Fig. 1g, h) similar to the dark-field TEM image and it can be also reflected in the lattice mapping with much better signal to noise ratio (Fig. 1i) as the polarization is highly correlated with the lattice. Therefore, the evolution of rotation direction for each vortex during in situ biasing can be reliably tracked by multiple parameters (e.g., dark field TEM image, out-of-plane polarization mapping, and lattice c mapping) in decent atomically resolved images.

**Phase-field simulation of chirality change.** Phase-field simulation is firstly performed to reveal the evolution of chirality in vortex under electrical fields. The clockwise and counterclockwise vortices shown in Fig. 2a have mixed axial polarization, in which the overlying red and blue colors denote the directions. Applying an out-of-plane electrical field leads to the formation of uniform c-domain and annihilation of chirality (Fig. 2b). With the removal of electrical fields, the in-plane polarization component ($P_{[010]}$) redistributes concomitant with the gradual recovery of vortex feature (Fig. 2c-e). Such reformation can finally result in the reversal of axial polarization (Fig. 2f), leading to the chirality switching. Furthermore, during the erasure and recover process, the vortex positions can shift along the in-plane direction as shown in Fig. 2f (which indeed is observed experimentally in Supplementary Fig. 4). In this case, we can also expect chirality flip if the vortex is moved in one-cell distance (~ 5 nm) when the clockwise vortex takes the position of its counterclockwise neighbor's since they have opposite chirality.

**Experimental observation of rotation change.** Experimentally, the electrical field is applied to the vortex between the probe and SRO bottom electrode as schematically illustrated in Supplementary Fig. 5. With the bias increasing to 9 V, the alternating bright and dark contrast (Fig. 2g) becomes uniform underneath the probe (Fig. 2h, Movie S1), indicating realignment of the electric dipoles in the vortex (i.e. polarization switching). The fully switched region with uniform contrast highlighted by the red dashed outlines is c-domain that is evidenced by the in situ electron diffraction experiment in the Supplementary Fig. 6. Close inspection (Supplementary Fig. 7) reveals that surrounding the fully switched region there is an intermediated phase zone where the pristine dark doted contrast became into tilted stripes, which indeed is a feature of *a*/c domains as we discuss details below. Removal of electric fields results in spontaneous back-switching to form the vortex array (Fig. 2i).

More importantly, we observe that in some regions the vortex rotation can be reversed during switching and back-switching events, which is expected to lead to chirality switching given unchanged axial-components. The subtraction of Fig. 2g and 2i acquired before and after electric bias results in obvious nonuniform contrast in the marked region (Fig. 2j), indicating the occurrence of rotation change, which is further confirmed by the enlarged view of the same region with the rotation sense denoted by the arrows (bottom row of Fig. 2g, i, j), while the rest region remains almost the same after removing the bias.

In order to determine the vortex position more accurately, we perform atomically resolved in-situ TEM experiment. Figure 3a, d are the atomically resolved HAADF-STEM images acquired before and after electric fields for the exactly same region that is tracked by a deliberately made mark (red circle) using electron beam (Methods). With the same label rule in Fig.1d and e, the rotation direction of the clockwise and counter-clockwise vortices are labeled by the curved arrows from the out-of-plane displacement distribution (Fig. 3b, e)[27] and out-of-plane lattice mapping (Fig. 3c, f). In the middle PTO layer, the rotation direction of vortex (labeled by blue curved arrows) is reversed after the switching events, while for the upper and lower PTO layer they keep unchanged. Such a rotation switching can change the local chirality as well as the direction of the toroidal moment[22] once the axial-component of polarization remains the same.

**The motion of vortex under electric field.** In fact, besides the evolution of rotation direction, the detailed switching pathways can be revealed. The transition process is recorded at atomic resolution and the polar feature is presented with the unit-cell mapping of the out-of-plane lattice. Application of a positive 6 V bias results in formation of striped *a*/c domains as shown in Fig. 4c, d, while the pristine unswitched region presents sinusoidal feature in Fig. 4a, b (see also Supplementary Fig. 8 and Supplementary Fig. 9). Before the destruction of the vortex core, the electric field that favors upward polarization and suppresses the downward one, is expected to drive the lateral movement of the vortex cores to adapt the expansion of the upward domain and shrinkage of the downward domain, forming the close-pair structure[13]. With increasing the bias to 12 V, the electric dipoles are fully aligned along the out-of-plane direction to form a pure upward domain in Fig. 4e, f. Similar transition is observed under negative bias (Supplementary Fig. 10-11). Such a two-stage phase transition from vortex to c-domain with intermediate $a$/c striped domain were predicted by the phase field simulations[13]. Our experimental observations are different than the previously reported non-contact long electric pulse bias induced switching event[15], in which the vortex would transform to dipolar wave instead of the $a$/c domain stripe. Furthermore, such intermediate $a$/c domain stripe was also not observed from the mechanical switching of the vortex[17], indicating the switching paths for these topological structures also largely depend on external stimuli.

**Discussion**

The underlying mechanism of the change of vortex rotation can be understood as follows. On one hand, the occurrence of intermediated $a$/c phase may give rise to the change of the core positions. In fact, in Fig. 4d, the core distance highlighted by the arrow is reduced from 12 to 8-unit cells under electric field in order to adapt the realignment of electric dipoles (i.e., expansion of upward domain and shrinkage of downward domain). Previously, the bias driven lateral movement of the vortex cores has been theoretically predicted to reduce the electrostatic energy[13]. In this case, the back-switching starting from these intermediate states with either vortex close-pair or $a$/c domain stripes may not precisely backtrack due to the possible disturbance from random fields[28, 29] and thus in the end the rotation of vortices should be different than the original states, leading to the chirality change. On the other hand, the distributions of strain and static fields are supposed to be uniform laterally once the polarization in vortex is switched to mono c-domain. Thus, there is no preference for the formation of the vortices during the subsequently spontaneous back-switching and the new core positions of newly generated vortex don't have to be the same with the previous ones.

Although our simulations and experiments demonstrated that the rotation reversal and position change of vortex is possible, we have to admit that at current stage neither the rotation reversal nor position shift can be well controlled as during the spontaneous back-switching the nucleation of new vortex can be easily disturbed by the random fields. Precise control of the chirality of single vortex is indeed possible, but require more sophisticated devices. Application of an electric field with both out-of-plane (along [001]) and in-plane components (along [010]) can tune the axial polarization to initialize an ordered chirality in the vortex array (Supplementary Fig. 12a-d). Moreover, single vortex chirality can be realized through a localized electric bias (Supplementary Fig. 12e-h). Such a field may be realized in future through surface probe techniques or patterned with nanosized electrodes from cross-sectional view.

In summary, this work reveals the phase transition behavior of individual vortices at the atomic scale. The transition starts with vortex core motion and evolve to *a*/c domain then finally to c-domain with increasing bias. After turning off the bias, vortex array spontaneously recovers. During the back switching, the rotation of vortex can be different than the as-grown state, leading to chirality change. The controlled conversion between topological vortex and ferroelectric phase may advance the development of nanoelectronic devices with novel functionalities. Particularly, the demonstrated ability to electrically switch the rotation direction and move the core position of vortex provides a taking one step forward to control the chirality by using the simple external electric field and further use such an order parameter for applications.

**Acknowledgments**

This work was supported by the Program from Chinese Academy of Science (Grants Nos. ZDYZ2015-1 and XDB33030200); the National Key R&D Program of China (Grants No. 2016YFA0300804); the National Natural Science Foundation of China (Grants Nos. 11974023, 21773303, 11875229, 51872251, 51991340, and 51991344); the Key R&D Program of Guangdong Province (Grants Nos. 2018B030327001, 2018B010109009, and 2019B010931001); Bureau of Industry and Information Technology of Shenzhen (Grants No. 201901161512); Beijing Excellent Talents Training Support (Grants No. 2017000026833ZK11); and the "2011 Program" Peking-Tsinghua-IOP Collaborative Innovation Center for Quantum Matter.

**Author contributions**

P.G., C.B.T., J.W. and X.D.B. conceived the idea; C.B.T. grew the samples with the help of J.B.W., X.P.O., and X.L.Z.. Z.X.J. simulated and predicted the experimental results directed by J.W.; P.C. performed the electron microscopy experiments and data analysis assisted by Y.W.S, X.M.L, L.L, L.F.W., X.M.L., W.L.W., K.Q., R.X.Z., N.L., and K.H.L under the direction of P.G.and X.D.B.; Z.X. contributed in situ TEM holders; P.C. drafted the manuscript. X.D.B. supervised the entire project.

**Competing interests**

The authors declare that they have no competing interests.

**Data availability**

The data that support the plots within this paper and other findings of this study are available from the corresponding author upon reasonable request.

**Materials and Methods**

**Thin Film Growth**

The bottom electrode layer of ~10 nm $SrRuO_3$ was grown on $(110)o$-$DyScO_3$

substrates followed by the growth of the $(PbTiO_3)_n/(SrTiO_3)_n$ superlattice films by a pulsed laser deposition (PLD) system (PVD-5000) using a KrF excimer laser (λ=248 nm). During the growth of SrRuO$_3$, the substrate temperature was kept at 690 ℃ with an oxygen pressure of 80 mTorr and a laser energy of 390 mJ pulse$^{-1}$, and when the substrate temperature was cooled down to 600 ℃, the n uc-thickness PbTiO$_3$ and SrTiO$_3$ layers were alternately deposited at an oxygen pressure of 200 mTorr and a laser energy of 340 mJ pulse$^{-1}$. Thicknesses of the SrRuO$_3$, PbTiO$_3$ and SrTiO$_3$ layers were held by controlling the growth time under a pulse repetition rate of 10 Hz. Ceramic targets Pb$_{1.1}$TiO$_3$ with a 10 mol% excessive amount of lead was used to compensate the evaporation loss of Pb during the growth of PbTiO$_3$ layer. After growth of the 15-$(PbTiO_3)_n/(SrTiO_3)_n$ cycles, the samples were cooled down to room temperature at 10 ℃/min in 200 mTorr oxygen pressure.

**TEM Sample Preparation**

TEM cross-sectional samples were polished to about 30 micrometers made by conventional mechanical methods and then transferred to ion-beam milling process in Gatan PIPS 695 system with an acceleration voltage of 3 kV for a quick milling followed by the final 0.1 kV to reduce the damage.

**STEM Characterization and analysis**

Atomic resolution HAADF images were obtained in a double aberration-corrected JEOL Grand ARM 300 CFEG with collection semi-angle [54, 220] mard and convergence semi-angle 18 mrad. STEM images were filtered by a HRTEM-filter plugin to remove the zero-frequency and high-frequency noise. The polarization map was done by calculating the offset between A site and B site using a Matlab code based two-dimensional Gaussian algorithm. Geometric phase analysis images were obtained by a free FRWRtools plugin for the Digital Micrograph software.

**In situ (S)TEM.**

In situ TEM experiments were carried out using a customized holder from ZepTools Technology Company on JEOL ARM 300F, allowing us to get atomic resolution even under the applications of electric fields because of the good stability of the holder. Electric bias was applied to the PTO/STO films with the scanning tungsten tip acting as a top electrode and the conductive SRO layer serving as the bottom electrode. In situ experiment was carried out in TEM mode for the dark-field and electron diffraction. Dark-field images were acquired under two-beam condition with $(002)_{pc}$ selected for the optimal image condition. The atomic HAADF-STEM images were acquired under the application of electric fields with a dwell time of 4 μs per pixel. In STEM mode, whether the electric fields are applied to the samples can be judged from the current values. The markers were made by focusing the electron probe on a specific location for a few minutes.

**Phase Field Modeling**

In order to study the chirality of the polar vortices and its evolution under an electric field, a phase field model is employed to simulate the polarization evolution in ferroelectric superlattice. In the phase field model, the free energy $F$ is the sum of Landau free energy, domain wall energy, elastic energy and electric energy. The total free energy of the PTO/STO ferroelectric superlattice has the following form[30]:

$$F = \int [\alpha_i P_i^2 + \alpha_{ij} P_i^2 P_j^2 + \alpha_{ijk} P_i^2 P_j^2 P_k^2 + \frac{1}{2} g_{ijkl} \left( \frac{\partial P_i}{\partial x_j} \frac{\partial P_k}{\partial x_l} \right) + \frac{1}{2} c_{ijkl} \varepsilon_{ij} \varepsilon_{kl}$$

$$- q_{ijkl} \varepsilon_{ij} P_k P_l - \frac{1}{2} \varepsilon_0 \varepsilon_r E_i E_i - E_i P_i ] dV \quad (1)$$

in which $\alpha_i$, $\alpha_{ij}$ and $\alpha_{ijk}$ denote the Landau coefficients, $g_{ijkl}$ is the gradient coefficient, $c_{ijkl}$ is the elastic constant, $q_{ijkl}$ is the electrostrictive coefficient and $\varepsilon_0$ is the vacuum permittivity. All values of the above coefficients at room temperature can be found in the previous papers[31-33]. As for $\varepsilon_r$, the relative dielectric constant of the background material, is set to 66 and 300 for PTO and STO respectively. $P_i$, $\varepsilon_{ij}$ and $E_i$ represent the polarization, strain and electric field, respectively. The repeating subscripts in equation (1) imply summation over the Cartesian coordinate components $x_i$ ($i = 1, 2$ and $3$).

In this study, the temporal evolution of the domain structure is obtained by solving the time-dependent Ginzburg-Landau (TDGL) equations[34]:

$$\frac{\partial P_i(\mathbf{r}, t)}{\partial t} = -L \frac{\delta F}{\delta P_i(\mathbf{r}, t)} \quad (2)$$

where $L$, $\mathbf{r}$, and $t$ denote the kinetic coefficient, spatial position vector, and time respectively. The TDGL equations choose the polarization vector $\mathbf{P} = (P_1, P_2, P_3)$ as the order parameter, and the domain structures of ferroelectric superlattices are described by polarization spatial distribution.

In addition, both the mechanical equilibrium equation:

$$\sigma_{ij,j} = \frac{\partial}{\partial x_j}\left(\frac{\partial f}{\partial \varepsilon_{ij}}\right) = 0 \tag{3}$$

and the Maxwell's equation:

$$D_{i,i} = -\frac{\partial}{\partial x_i}\left(\frac{\partial f}{\partial E_i}\right) = 0 \tag{4}$$

are satisfied at the same time, where $\sigma_{ij}$ and the $D_i$ are the stress and electric displacement components, respectively. In this work, the semi-implicit Fourier-spectral method is employed to numerically solve the partial differential equations[35].

The thickness of each layer in (PbTiO$_3$)$_{10}$/(SrTiO$_3$)$_{10}$ superlattices is set to 4 nm and the length and width of the model is set to 40 nm. A 3D mesh of $100 \times 100 \times 60$ grids is used and each grid is set to 0.4 nm to simulate the lattice size of the material approximately. Periodic boundary conditions are applied on the stack direction and in-plane dimensions. Small random fluctuation (<$0.01P_0$, where $P_0 = 0.757 \text{ Cm}^{-2}$ is the spontaneous polarization of PTO at room temperature) is used as the initial values of polarization to initiate the polarization evolution. To be consistent with the experiments, the same biaxial misfit strain of -0.5% is applied in the simulations. The gradient coefficients are assumed to be isotropic with $g_{11} = 1.038 \times 10^{-10}$ and $g_{44} = 0.519 \times 10^{-10}$ m$^4$NC$^{-2}$. What's more, the external electric field used to regulate chiral vortices is uniformly applied.

# Figures

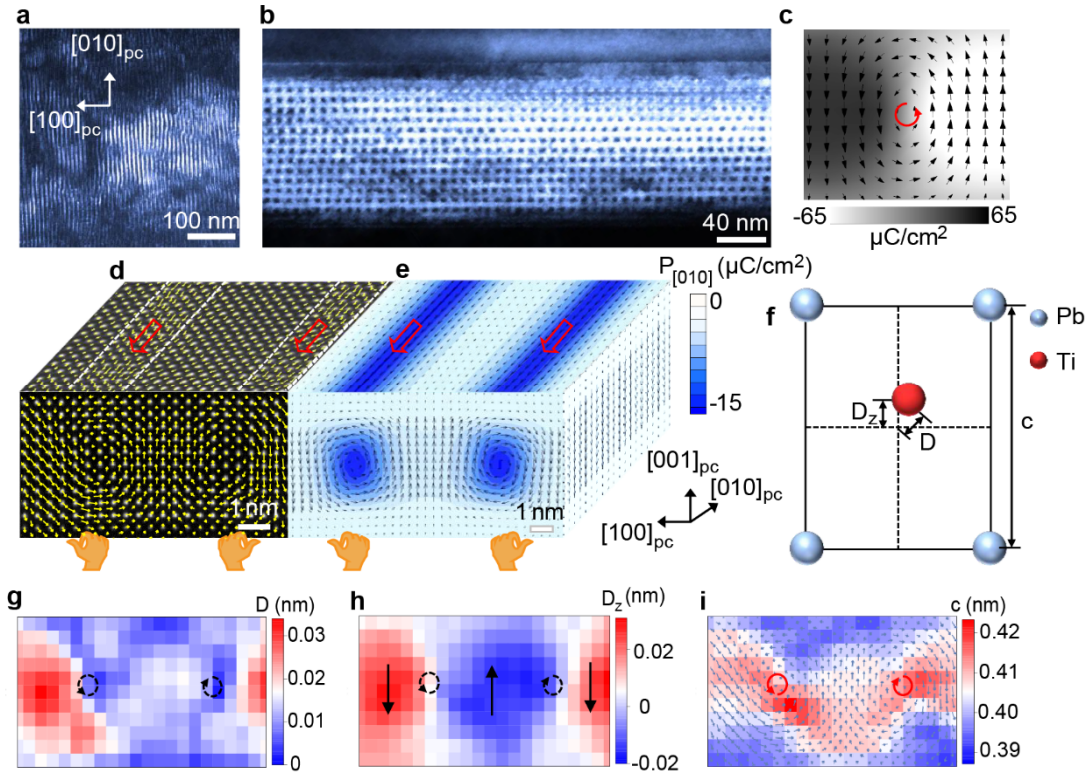

**Fig. 1 | Characterization of vortex chirality in PbTiO$_3$/SrTiO$_3$ superlattices. a** Plane-view dark-field image, indicating the long-range in-plane ordering. **b** A cross-sectional dark field image under two beam condition (vector g=002$_{pc}$, pc denotes pseduocubic indices) with alternative bright and dark contrast corresponding to the clockwise and counterclockwise vortex pairs. **c** The distribution of out-of-plane polarization simulated by phase-field simulations with the bright and dark contrast resembling the contrast of dark-field image. The vortex core locates in the boundary of the bright and dark area. **d** Atomically resolved HAADF-STEM image with the yellow displacement vectors exhibiting a vortex pair. The image overlaid above is a HAADF-STEM image showing a parallel axial polarization for a vortex pair. **e** The configuration of vortex reproduced by phase-field simulation, in good agreement with experimental results. As the chirality of vortex is determined by both the vortex rotation and axial polarization, an opposite chirality (indicated by the orange hand label) is thus derived

by curling the fingers along the rotation and thumbs pointing to the direction of axial polarization. **f** A schematic of the perovskite PbTiO$_3$ structure with the notation $D_z$, D and c representing the [001] component of displacement, total displacement and c lattice constant, respectively. **g-i** The total polarization, out-of-plane polarization distribution and out-of-plane lattice mapping of the vortex in **d**, respectively. The core positions and rotation directions can be identified from these maps.

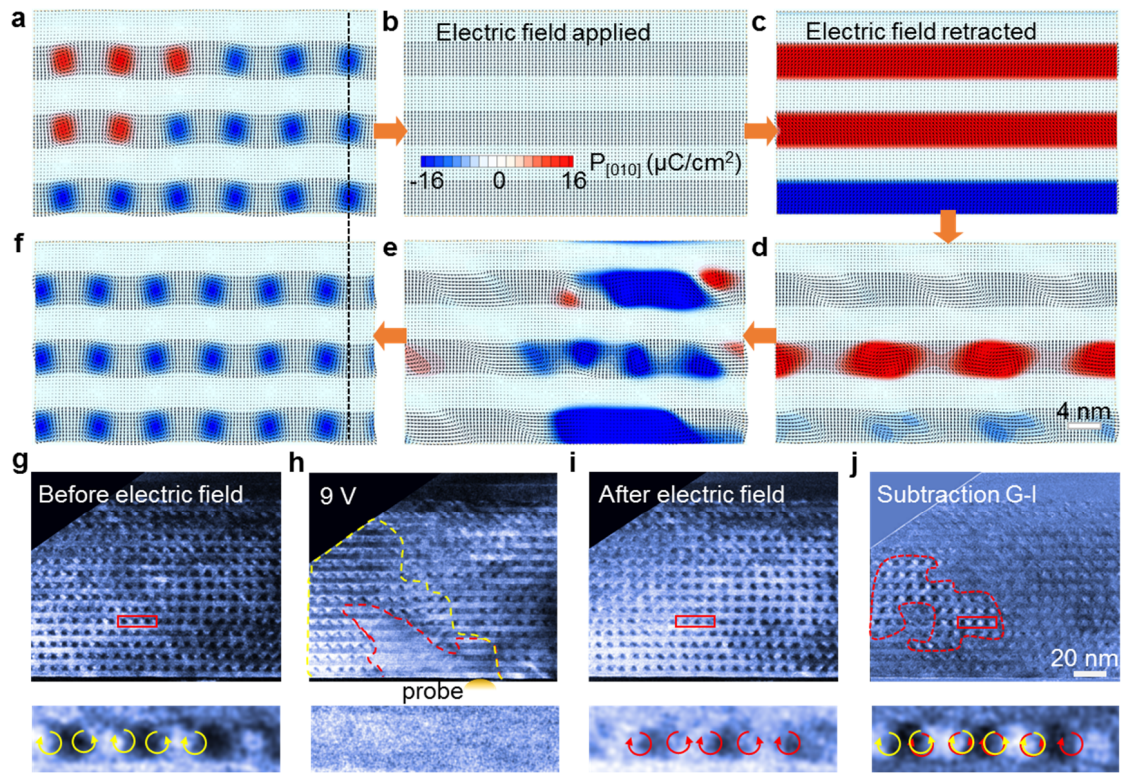

**Fig. 2 | Chirality switching of vortex by electric field. a** The pristine vortex array with the blue and red colors indicating the direction of axial component. The handedness thus can be determined for individual vortex. **b** Uniform c-domain formed under an out-of-plane electric field of $E_{[001]}=261.2$ kV/cm. **c-f** The recovery of vortex with the removal of electric fields. The redistribution of the polarization eventually give rise to the shift of vortex positions and the reversal of axial polarization, which can cause a chirality change. **g-i** The dark-field image acquired before, under and after electric fields, respectively. Application of a 9 V bias leads to the vortex transition into uniform contrast (red dashed line) and tilt stripes (yellow dashed line) under the probe. **j** The subtraction of image **g** and **i**. Note most part of the image shows no contrast except for the marked region (red dashed line). The bottom row shows the rotation reversal of vortex before and after electric field from the same region (red boxes) as the rotation direction is indicated by yellow and red arrows.

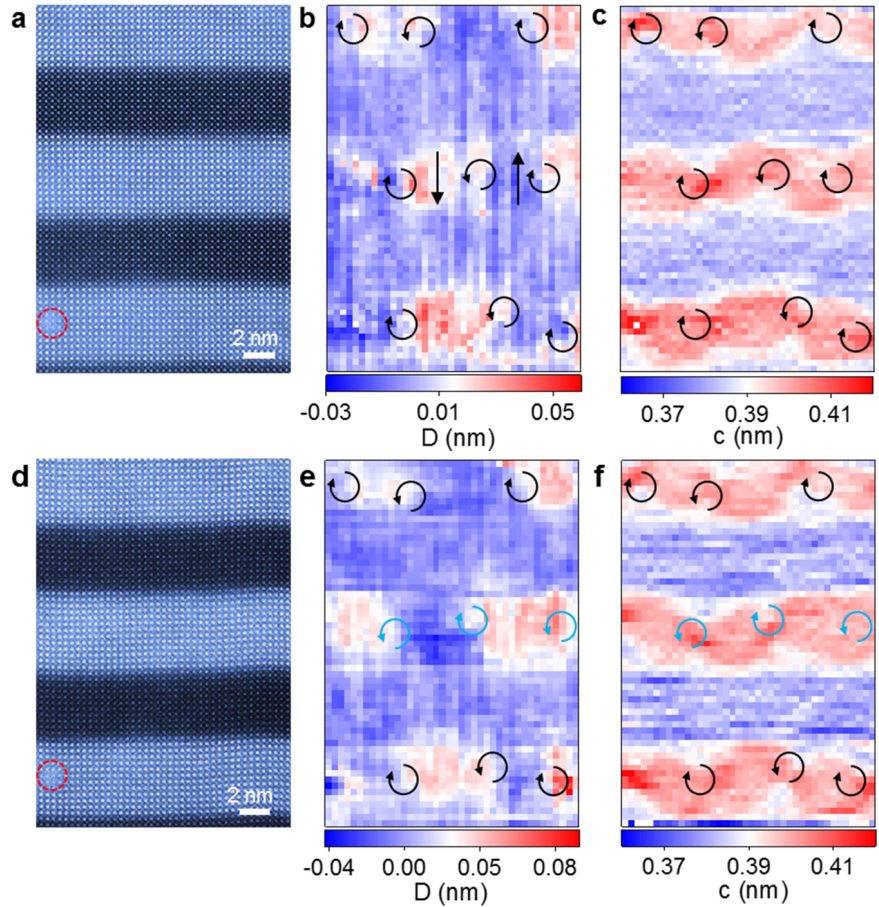

**Fig. 3 | Vortex rotation reversal. a** A HAADF-STEM image of the pristine vortex with a mark (denoted by the red circle) deliberately made by the focused electron beam. **b, c** The corresponding out-of-plane displacement (denoted as D) and c lattice mapping of vortex array. **d** A HAADF-STEM image and **e, f** the corresponding out-of-plane displacement and c lattice mapping after removal of bias. The positions of vortex cores and rotation direction can be determined by the out-of-plane displacement mapping, although the noise is relatively large due to the lager vibration of in-situ environment. While the positions of vortex in the upper and bottom PTO layers maintain unaltered, the vortex in middle PTO layer shows a rotation reversal.

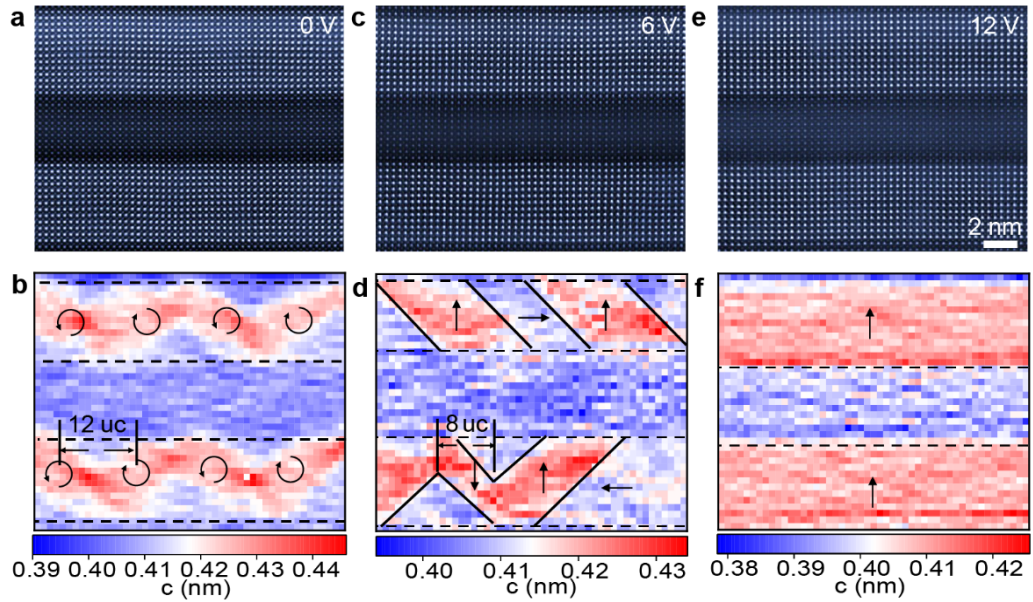

**Fig. 4 | Unit-cell scale mapping of the lattice constant. a** HAADF-STEM image recorded from the pristine region with vortex array. **b** The corresponding lattice c mapping. The distance between the two cores are typically 12 unit cells. **c** A HAADF-STEM image acquired from the region with intermediate phase at 6 V and **d** the corresponding lattice c mapping. The sinusoidal wave changes into tilted stripes, indicating formation of *a*/c domains. The c-domain with upward polarization is growing and c-domain with downward polarization is shrinking. The distance between the two highlighted cores are reduced to 8 unit cells. **e** A HAADF-STEM image acquired from fully switched region and **f** the corresponding lattice c mapping, confirming the final phase is c-domain with upward polarization.

# Supplementary Information For:

# Manipulation of polar vortex chirality in oxide superlattices

Pan Chen, Congbing Tan, Zhexin Jiang, Peng Gao, Yuanwei Sun, Xiaomei Li, Ruixue Zhu, Lei Liao, Xu Hou, Lifen Wang, Ke Qu, Ning Li, Xiaomin Li, Zhi Xu, Kaihui Liu, Wenlong Wang, Jinbin Wang, Xiaoping Ouyang, Xiangli Zhong, Jie Wang and Xuedong Bai

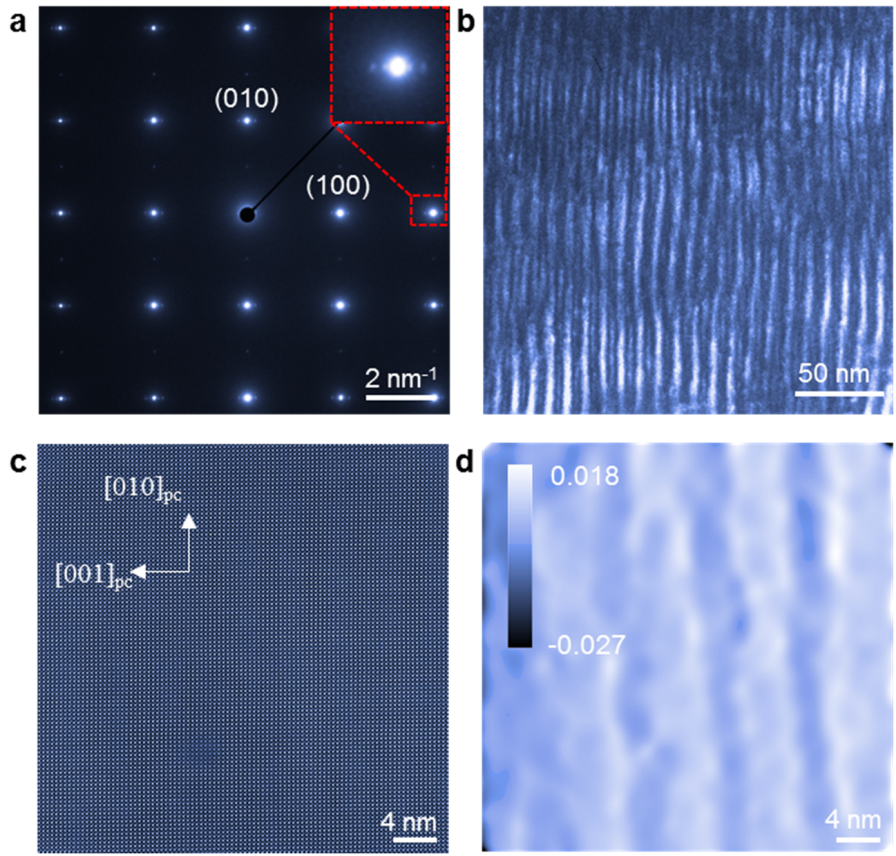

**Supplementary Fig. 1 | In-plane long-range ordering of PTO/STO superlattice. a** The selected aperture electron diffraction pattern (SAED) and **b** the plane-view dark-field image, indicating the in-plane ordering. The red rectangle is the g-vector used in **b**, which is magnified (inset). **c, d** the HAADF-STEM image and corresponding GPA component $\varepsilon_{xx}$, showing the tube like contrast.

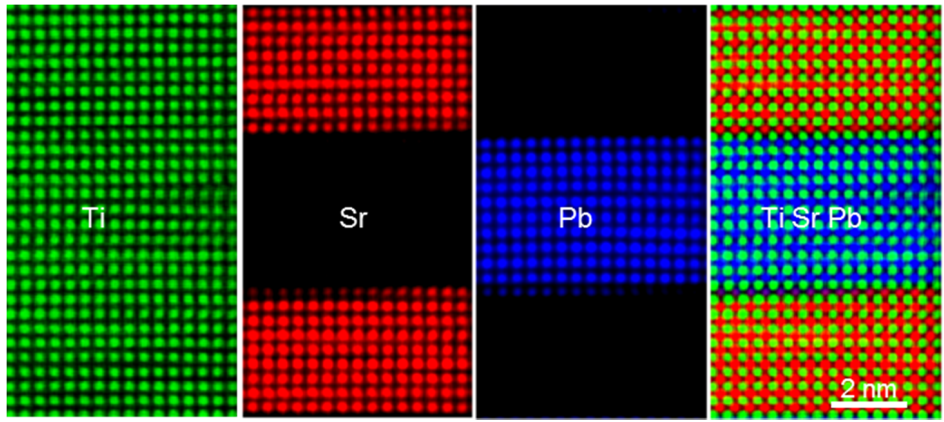

**Supplementary Fig. 2. | Characterization of PTO/STO superlattice.** EDS mapping of the cross-sectional PTO/STO superlattice film. The interface of the PTO and STO is sharp, indicating the good quality of the film.

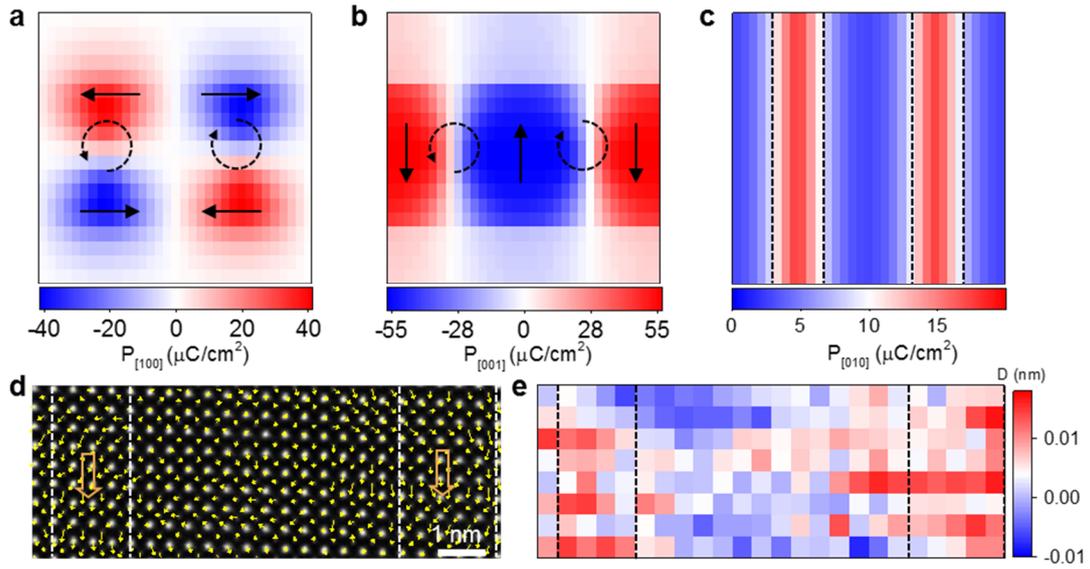

**Supplementary Fig. 3 | Quantitative analysis of the in-plane polarization. a-c** The quantitative distribution of the polarization along [100], [001] and [010], respectively, from phase-field simulation. **d** A HAADF-STEM image showing a parallel axial polarization for a vortex pair, similar to that of Fig. 1c. The image is overlaid by three HAADF-STEM image and then filtered in order to reduce the noise for accurate quantitative analysis. **e** The experimental displacement distribution along [010], showing a rather small value of polarization.

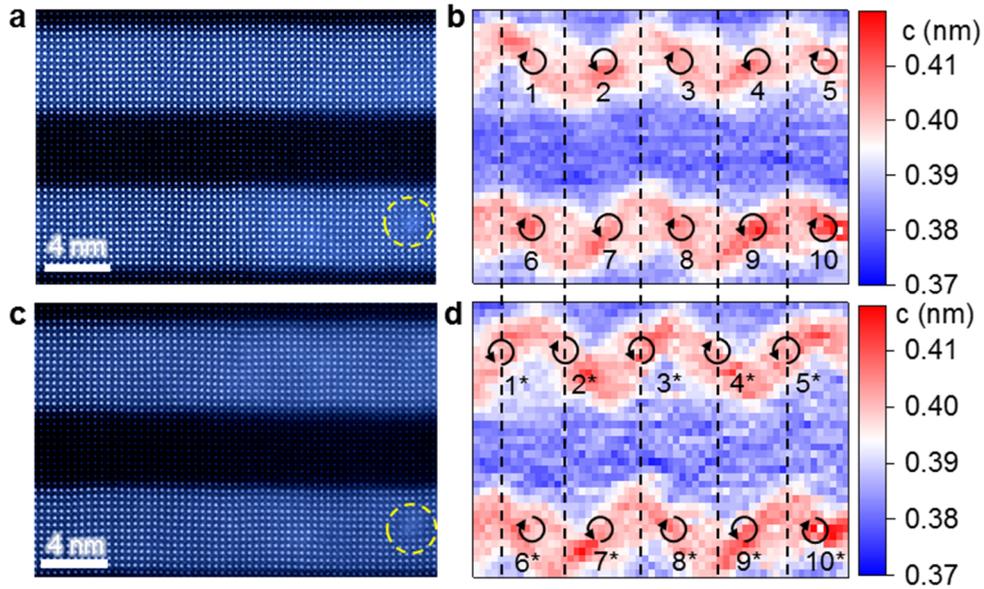

**Supplementary Fig. 4 | Vortex core position change. a** A HAADF-STEM image of the pristine vortex with a mark (denoted by the yellow circle) deliberately made by the focused electron beam. **b** The corresponding c lattice mapping of vortex array. **c** A HAADF-STEM image and **d** the corresponding c lattice mapping after application and subsequent removal of bias. The dashed lines highlight the features of sinusoidal wave for comparison. While the positions of vortex 6-10 nearly maintain unaltered at the bottom PTO layer, the vortex 1-5 shows distinguished feature compared to the vortex1*-5* after electric bias, indicating the occurrence of core position change.

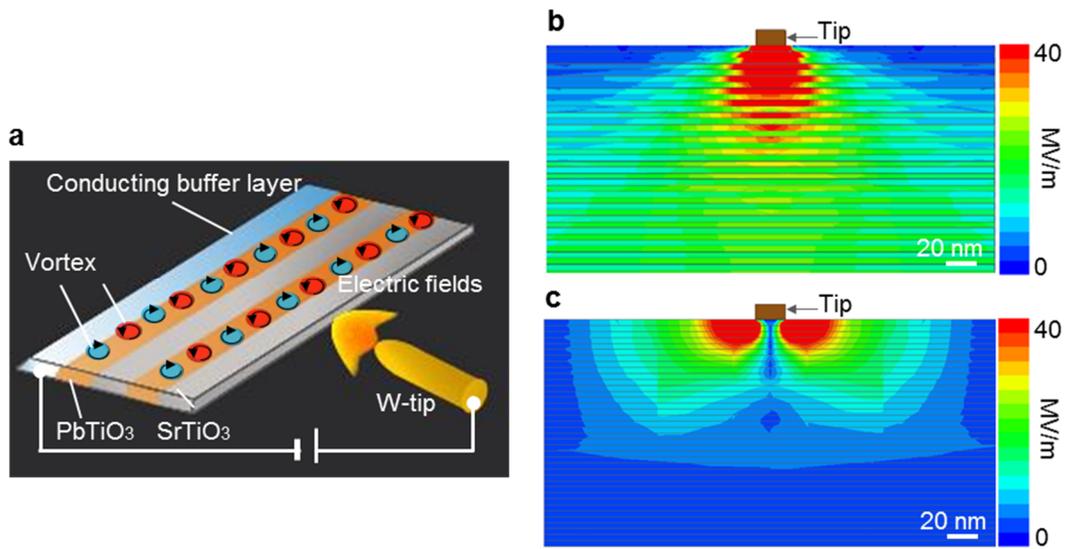

**Supplementary Fig. 5 | A schematic of in situ experimental setup and simulated distribution of electric field in the PTO/STO superlattice film. a** The in situ setup. **b, c** The asymmetric geometry of the electrodes induced distribution of out-of-plane and in-plane electric field component in the PTO/STO superlattice film simulated by finite element method.

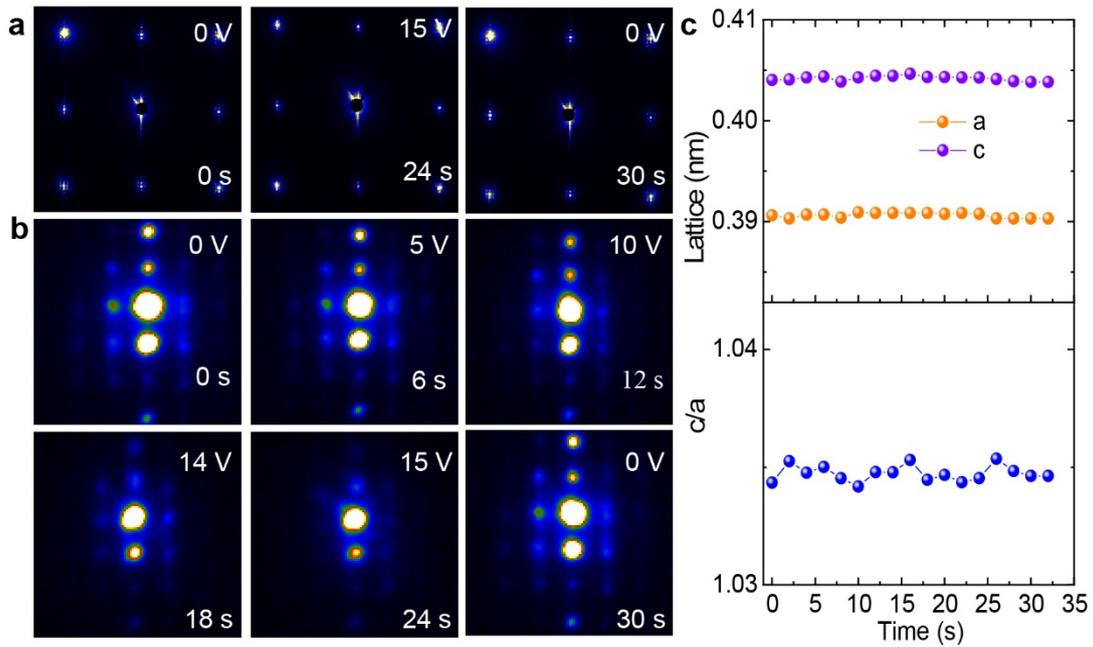

**Supplementary Fig. 6 | Electron diffraction tracking the structural evolution of vortex under positive electric fields. a** SAED images acquired before, during and after positive electric fields. **b** Close inspection of the $(001)_{pc}$ spots under the positive bias showing the vortex transition as the additional spots along in-plane direction, which correspond to the period of the vortex arrays, gradually disappeared. **c** Lattice parameters and the $c/a$ ratio versus time, indicating the out-of-plane polarization after switching since the ratio of out-of-plane to in-plane keeps larger than 1.

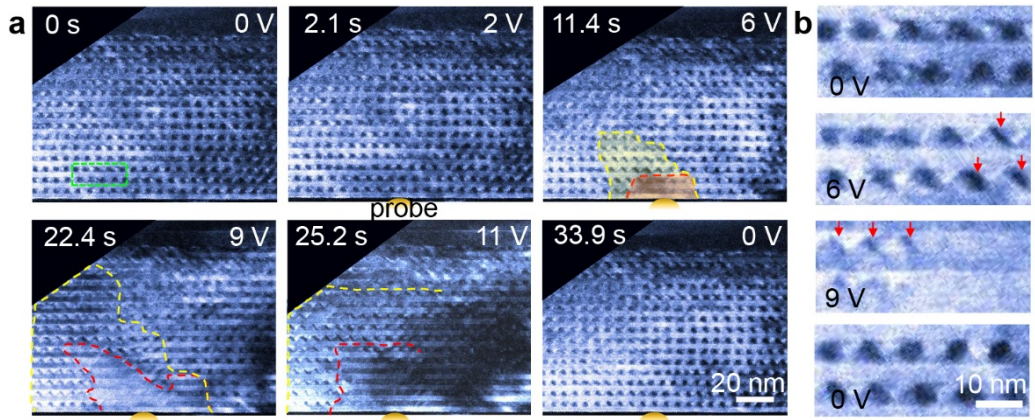

**Fig. S7 | Electrically driven transition of polar vortex. a** Consecutive TEM dark field images showing bias-induced transition from vortices to regular ferroelectric states. The red outline denotes the switched region of c-domain and the yellow dashed outline highlights the region of intermediate phase. At 25.2s, the outlines are not clear enough to be labeled in the right-side region. **b** Enlarged view of the vortex transition process from the selected region (green rectangle) in **a**. The arrows indicated the tilted stripes are intermediate phase with *a*/c stripe domains.

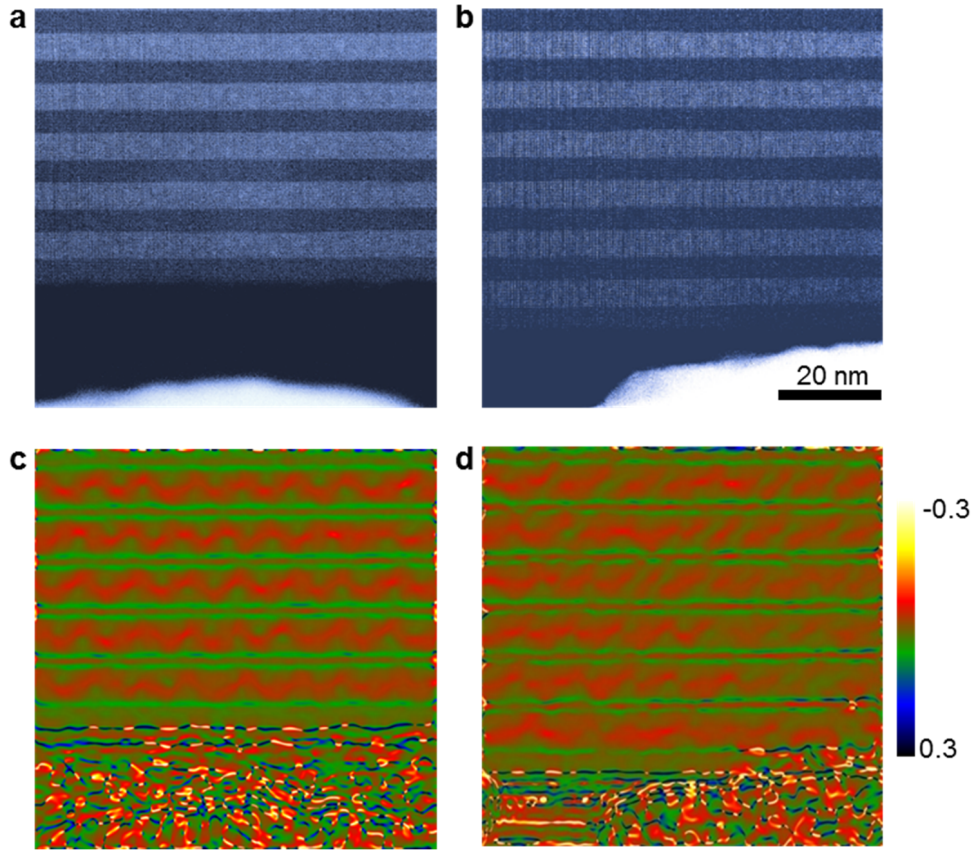

**Supplementary Fig. 8 | Confirmation of intermediated a/c domain under electric bias. a, b** HAADF-STEM image acquired before and under electric fields. **c, d** The corresponding GPA images with the sinusoidal wave contrast breaking into stripes, indicating the vortex become a/c domain.

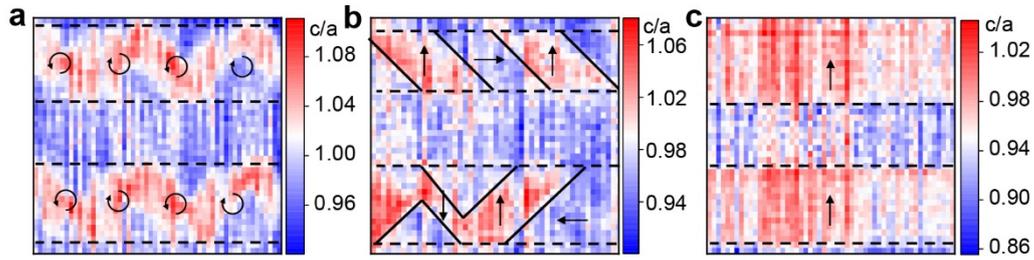

**Supplementary Fig. 9 | Unit-cell scale mapping of the lattice constant ratio. a-c** the c/a ratio mapping acquired from Fig. 4**a, c, e** in the main text. Although noise is relatively higher, the c/a ratio mapping shows the same trend as the lattice c mapping in Fig. 4**b, d, f,** validating a/c domain mediated phase transition from vortex to c domain. The noise mainly comes from the scanning noise, which induces large artificial fluctuation in the lattice a measurement.

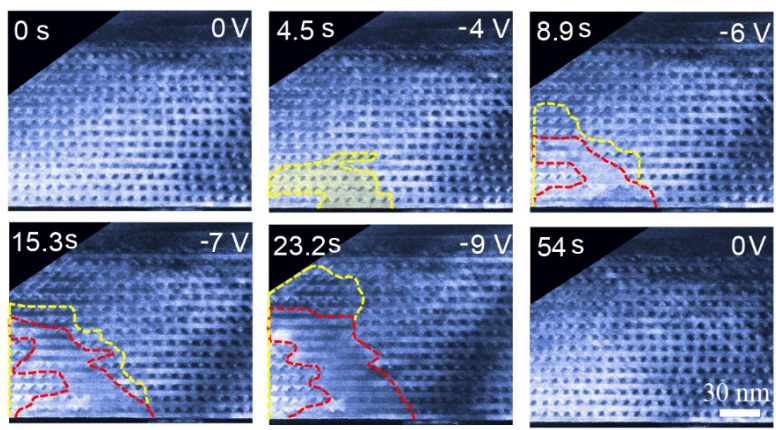

**Supplementary Fig. 10 | Dark field image showing the structural transition of vortex under negative bias.** Consecutive TEM dark field images showing bias-induced vortex transition as the alternative dark and bright contrast become uniform (intermediated transition area and uniform transition area indicated by the yellow lines and red lines, respectively).

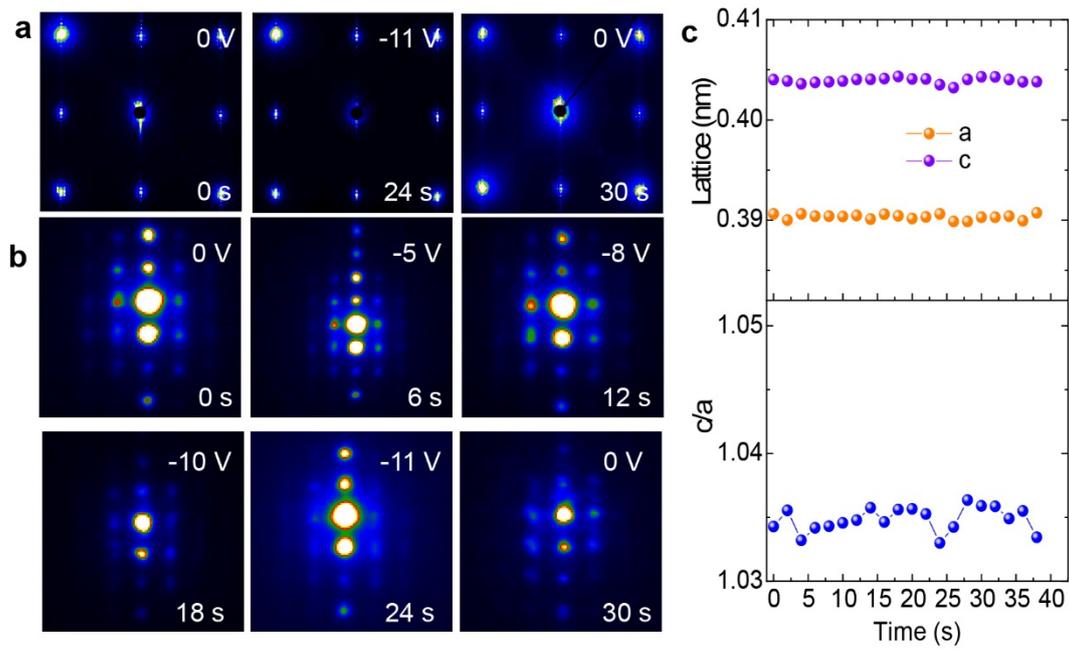

**Supplementary Fig. 11 | Electron diffraction tracking the structural evolution of vortex under negative electric fields. a** SAED images acquired before, during and after negative electric fields. **b** Close inspection of the $(001)_{pc}$ spots under the negative bias showing the vortex transition as the additional spots along in-plane direction gradually dimmed. **c** Lattice parameters and the c/a ratio versus time, indicating the out-of-plane polarization since the c/a ratio keep bigger than 1.

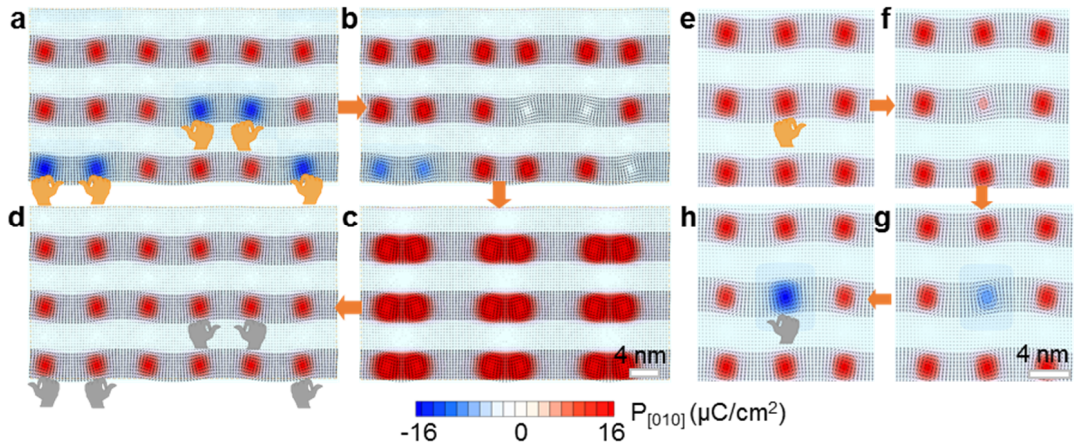

**Supplementary Fig. 12 | Control of vortex chirality. a** The pristine vortex array with the blue and red colors indicating the direction of axial component. **b-d** The evolution of vortex under an electric field with in-plane $E_{[010]}$=15.67 kV/cm and out-of-plane $E_{[001]}$=156.72 kV/cm. The direction of axial polarization become uniform, resulting in an ordered chirality for the vortex array. **e-h** The evolution of vortex under in-plane ([010] direction) electric field localized in a single vortex region ($E_{[010]}$=15.67 kV/cm). The direction of axial polarization can be switched, leading to a chirality reversal for single vortex.

**Movie S1.**

An *in situ* TEM dark-field movie showing the evolution of vortices under positive bias. The alternating bright and dark contrast shrank into stripes and then became uniform with the increasing bias. The video was recorded at 0.15 s/frame and played at 20 frame/s.